\documentclass[12pt,3p,times]{elsarticle}

\usepackage{amssymb}

\usepackage{amsmath}
\usepackage{bm}
\usepackage{hyperref}
\usepackage[hyperref]{xcolor}
\usepackage{subcaption}

\usepackage{lineno}

\usepackage[figuresright]{rotating}

\date{January 3, 2021}

\begin{document}

\begin{frontmatter}

\title{Quantum Fokker-Planck Modeling of Degenerate Electrons}

\author{Hai P. Le\corref{cor1}\fnref{label2}}
\ead{hle@llnl.gov}
\cortext[cor1]{Corresponding author.}

\address{Lawrence Livermore National Laboratory, Livermore, California 94551, USA}

\begin{abstract}
{An implicit and conservative numerical scheme is proposed for the isotropic quantum Fokker-Planck equation describing the evolution of degenerate electrons subject to elastic collisions with other electrons and ions}. The electron-ion and electron-electron collision operators are discretized using a discontinuous Galerkin method, and the electron energy distribution is updated by an implicit time integration method. The numerical scheme is designed to satisfy all conservation laws exactly. Numerical tests and comparisons with other modeling approaches are shown to demonstrate the accuracy and conservation properties of the proposed method.
\end{abstract}


\end{frontmatter}

\section{Introduction}
\label{sec:intro}
Coulomb collisions play a central role in plasma physics as the main driving mechanism for temperature equilibration and transport phenomena. They are typically modeled by a Fokker-Planck (FP) collision operator with the underlying assumptions that the particles are non-degenerate, weakly coupled and that small angle scattering dominates \cite{shkarofsky_particle_1966}. This approach works well for low density, high temperature `classical' plasmas. At higher densities and lower temperatures, strong Coulomb coupling and quantum effects becomes significant and the classical treatment is no longer applicable. Quantum kinetic theory for dense plasmas has developed over the last couple decades, but the numerical solution of these equations remains a challenge due to the complex mathematical structure involved \cite{kremp_quantum_2005,bonitz_quantum_2016}.

{In the present work, we focus on developing numerical solutions for an isotropic quantum Fokker-Planck (qFP) model of degenerate electrons. The electron distribution function is assumed to be spatially homogeneous and isotropic, such that the qFP model can be formulated in terms of the electron energy distribution function}. The qFP model is a direct extension of the classical model, in which the effects of quantum degeneracy are taken into account  \cite{danielewicz_nonrelativistic_1980,brown_transport_1997,daligault_quantum_2016}, including the Pauli exclusion principle, which forbids transitions into occupied states. The qFP equation can be derived from the grazing collision limit of the quantum Boltzmann equation (also known as the Boltzmann-Uehling-Uhlenbeck equation) \cite{uehling_transport_1933}. The domain of applicability for the qFP equation was examined by Daligault \cite{daligault_crossover_2017}. It can be applied to electron collisions in high energy density physics experiments, where the electrons remain weakly coupled over a wide range of temperature. In addition, since the formulation of the qFP equation is very similar to its classical version, existing numerical methods developed for the classical FP equation can be readily adapted \cite{chang_practical_1970,epperlein_implicit_1994,pareschi_fast_2000,taitano_adaptive_2016,le_conservative_2017}.

The classical FP model has been studied extensively in the context of plasma transport \cite{langdon_nonlinear_1980,epperlein_plasma_1986} and kinetic simulations based on the FP model are routinely used to study nonequilibrium systems \cite{epperlein_two-dimensional_1988,larroche_kinetic_1993,milder_evolution_2020}. The qFP model receives much less attention and very few studies discuss numerical methods for solving the qFP equation. Most of the studies dicussing about the qFP equation are in the context of plasma transport theory \cite{lampe_transport_1968,brown_transport_1999,daligault_collisional_2018}. We mention here those studies that discuss numerical solutions to qFP and/or quantum kinetic equations in general. Hu et al. \cite{hu_numerical_2012} developed an asymptotic preserving scheme for the qFP equation. In this work, the collision operator is discretized using the spectral method of Pareschi et al. \cite{pareschi_fast_2000}, originally designed for the classical FP equation. A similar scheme was also proposed for the quantum Boltzmann equation \cite{filbet_numerical_2012}. Kitamura \cite{kitamura_thermalization_2019} studied ultra-fast thermalization dynamics of primary and secondary electrons in metals by numerically solving the quantum Lenard-Balescu (LB) equation. A similar effort at solving the non-degenerate quantum LB equation was due to Scullard et al. \cite{scullard_numerical_2016,scullard_adaptive_2020}. We also note that Monte Carlo collision algorithms which simulate Coulomb collisions can be extended to include quantum degeneracy effects, and these provide an alternative to the qFP approach. \cite{turrell_monte_2013,borowik_modified_2017}

The numerical scheme presented here is built upon previous work on nonequilibrium electron kinetics modeling \cite{le_conservative_2017,le_influence_2019}. For simplicity, we consider the case of a uniform and spatially homogeneous plasma. We further assume that the electron distribution is isotropic and formulate the qFP operators for both electron-ion (ei) and electron-electron (ee) collisions. Degeneracy and strong coupling effects of ions are neglected, and the ion distribution is assumed to be Maxwellian. The resultant qFP equation is discretized and solved by a fully implicit Discontinuous Galerkin (DG) method. The discretization method is high-order, and can be applied to a non-uniform energy grid. The fully implicit time stepping scheme allows us to take time steps larger than the mean collision time. These features are attractive for problems with a large dynamical range of conditions. Other physics can be incorporated. For example, the same numerical formulation was applied to the case of inelastic collisions, e.g., excitation and ionization, which are modeled by Boltzmann collision operators \cite{le_conservative_2017}. The extension to include quantum degeneracy for inelastic collisions will be examined in future work. Transport effects due to spatial inhomogeneity can be incorporated by employing the full spherical Harmonic decomposition of the velocity distribution function \cite{shkarofsky_particle_1966,bell_fast_2006,thomas_review_2012}. This was demonstrated for the case of classical FP equation \cite{le_influence_2019}, and can be extended to the qFP equation in a straightforward manner.

The rest of the paper is organized as follows. In section 2, we summarize basic properties of degenerate electrons and the qFP model for ei and ee collisions. Section 3 describes the DG discretization the collision operators, its conservation properties and time integration method. Several numerical tests are presented in section 4 to validate the qFP model. Comparisons with Monte Carlo simulations and simplified model are also shown. Finally we make some concluding remarks in section 5. The formulation of the qFP equation is discussed in \ref{sec:qfp}.

\section{Quantum Fokker-Planck Model}
Before introducing the qFP model, we review some basic definitions and properties of degenerate electron distributions. Consider a spatially homogeneous and isotropic electron gas in a fully ionized plasma free of external fields. The electron energy distribution function (EEDF) evolves due to collisions among the electrons and with surrounding ions. Electron thermalization is primarily due to ee collisions, while ei collisions are responsible for temperature equilibration between the ions and electrons. In thermal equilibrium, the mean occupation number of a state of energy $\varepsilon$, denoted as $\tilde{f} (\varepsilon)$, follows a Fermi-Dirac distribution:
\begin{align}
\label{eq:fd}
\tilde{f}^\star (\varepsilon)  &= \frac{1}{1+e^{(\varepsilon-\mu)/T_e}},
\end{align}
where $\mu$ is the chemical potential and $T_e$ is the electron temperature. Here we use a superscript $(\star)$ to indicate thermal equilibrium condition. The EEDF $f(\varepsilon)$ is related to the mean occupation number $\tilde{f} (\varepsilon)$ as follows:
\begin{align}
f(\varepsilon) = 4\pi  \left( \frac{2m}{h^2} \right)^{3/2} \sqrt{\varepsilon} \tilde{f} (\varepsilon).
\end{align}
The total electron density $N_e$ and energy $E_e$ are given by the zeroth and first order energy moments of the distribution function:\begin{subequations}
\label{eq:moment1}
\begin{align}
N_e &= \int_0^\infty f (\varepsilon  ) \, d\varepsilon, \\
E_e &= \int_0^\infty \varepsilon f (\varepsilon ) \, d\varepsilon.
\end{align}
\end{subequations}
For a Fermi-Dirac distribution (\ref{eq:fd}), we have:
\begin{subequations}
\label{eq:moment2}
\begin{align}
N_e = \frac{4}{\sqrt{\pi}} \lambda^{-3} \mathcal{I}_{1/2} (\mu/T_e), \\
E_e = N_e T_e \frac{\mathcal{I}_{3/2} (\mu/T_e)}{\mathcal{I}_{1/2} (\mu/T_e)},
\end{align}
\end{subequations}
where $\lambda \equiv \frac{h}{\sqrt{2\pi m T_e}}$ is the electron thermal deBroglie wavelength and $\mathcal{I}_n (y) \equiv \int_0^\infty \frac{x^n}{1+e^{x-y}} \, dx$ is the Fermi-Dirac integral of order $n$. Unlike the classical Maxwell distribution, $T_e$ (and $\mu$) cannot be trivially expressed in terms of $N_e$ and $E_e$, i.e., inverting Eq.~(\ref{eq:moment2}) requires an iterative procedure or numerical fit \cite{antia_rational_1993}. Temperature and chemical potential are uniquely defined only when the electrons are in thermal equilibrium. In the case of a nonequilibrium distribution with fixed $N_e$ and $E_e$, Eq.~(\ref{eq:moment2}) can be used to determine the corresponding equilibrium distribution.

Two asymptotic limits can be associated with the Fermi-Dirac distribution: classical (high temperature) and strongly degenerate (low temperature). These can be characterized by the degeneracy parameter, defined as $T_e/E_F$ where $E_F \equiv \frac{\hbar^2}{2m} (3\pi^2 N_e)^{2/3}$ is the Fermi energy. In the classical limit ($T_e/E_F \gg 1$), $e^{\mu/T} \rightarrow \frac{1}{2}N_e \lambda^3$ and $E_e \rightarrow \frac{3}{2} N_e T_e$ and the distribution approaches the classical Maxwellian. For a strongly degenerate system ($T_e/E_F \ll 1$), $\mu \rightarrow E_F$ and $E_e \rightarrow \frac{3}{5} N_e E_F$. At the zero temperature limit, all energy states up to $E_F$ are fully occupied, i.e,. $\tilde{f} (\varepsilon) = 1$ for $\varepsilon \leq E_F$ and 0 otherwise.

In a nonequilibrium system, the time evolution of the EEDF subject to ei and ee collisions obeys the following qFP equation (see \ref{sec:qfp} for the mathematical formulation):
\begin{align}
\label{eq:qfp}
\frac{\partial f(\varepsilon,t)} {\partial t}  = \frac{\partial } {\partial \varepsilon} \left( J_{ei} + J_{ee}\right).
\end{align}
Both ei and ee collision operators are written as a divergence of particle fluxes in energy space. The flux due to ei collisions is written as:
\begin{subequations}
\label{eq:ei}
\begin{align}
J_{ei} &= \gamma_{ei} \left[ \frac{f}{\sqrt{\varepsilon}}  \left( 1 - \tilde{f} \right) + T_i \frac{\partial}{\partial \varepsilon} \left( \frac{f}{\sqrt{\varepsilon}}\right) \right], \\
\gamma_{ei} &= \frac{m}{M} \sqrt{\frac{2}{m}} 2 \pi N_e Z e^4 \ln \varLambda_{ei},
\end{align}
\end{subequations}
where $Z$ is the ion charge state, $M$ is the ion mass, and $T_i$ is the ion temperature. Eq.~(\ref{eq:ei}) assumes that the ions are non-degenerate and thermalized, i.e., their energy distribution is a Maxwellian distribution at $T_i$. It also assumes that the ions' drift velocities are negligible and their thermal velocities are much lower than those of the electrons, i.e., $\frac{T_i}{M} \ll \frac{T_e}{m}$. The particle flux due to ee collisions is:
\begin{subequations}
\label{eq:ee}
\begin{align}
\label{eq:ee_j}
J_{ee} &= \gamma \left[ L \frac{f}{\sqrt{\varepsilon}} (1- \tilde{f}) + K \frac{\partial}{\partial \varepsilon} \left( \frac{f}{\sqrt{\varepsilon}}\right)  \right], \\
\label{eq:ee_gamma}
\gamma &= \sqrt{\frac{2}{m}} \frac{2\pi e^4 \ln \varLambda_{ee}}{3}, \\
\label{eq:ee_l}
L &= 3 \int_0^\varepsilon f \, d\varepsilon',\\
\label{eq:ee_k}
K &= 2 \int_0^\varepsilon f (1-\tilde{f}) \varepsilon' \, d \varepsilon' + 2\varepsilon^{3/2} \int_\varepsilon^\infty f(1-\tilde{f}) \varepsilon'^{-1/2} \, d \varepsilon',
\end{align}
\end{subequations}
The two logarithmic terms $\ln \varLambda_{ei}$ and $\ln \varLambda_{ee}$ in Eqs. (\ref{eq:ei}) and (\ref{eq:ee}) arise from the integral of the Coulomb differential cross section over all impact parameters. It is well-known that this integral diverges at both the lower and upper limits, so cut-offs must be imposed. It is often assumed, in the case of a dense plasma, that the lower cut-off is determined by the thermal deBroglie wavelength, and the upper cut-off by Coulomb screening. However, at high enough density, the interparticle separation can be comparable or greater than the screening length, and therefore is also a suitable choice for the upper cut-off. Different choices of these cut-offs result in different formula for $\ln \varLambda$ (see \cite{daligault_quantum_2016,gericke_dense_2002} and the references therein). For the purpose of this work, we will assume that $\ln \varLambda_{ei}$ and $\ln \varLambda_{ee}$ are constant and focus on the numerical discretization of the collision operators. The qFP Eq.~(\ref{eq:qfp}) must be solved subjected to the boundary condition that the particle fluxes $J_{ei}$ and $J_{ee}$ vanish at $\varepsilon = 0$ and $\infty$. The blocking factors $(1-\tilde{f})$ in (\ref{eq:ei}) and (\ref{eq:ee}) are direct consequences of the Pauli exclusion principle. These blocking factors increase the complexity and non-linearity of the kinetic equation, i.e., the ei flux is at most quadratic and the ee flux cubic in $f$. Similar to the classical case, the qFP collision operators (\ref{eq:ei}) and (\ref{eq:ee}) conserve density and energy. The flux divergence form of the collision operators and imposed boundary conditions guarantee conservation of total electron density for both ei and ee collisions. As a result, most finite volume discretizations easily satisfy this condition. It is more difficult to achieve energy conservation since it involves a high-order moment quantity. For ei collisions, the statement of energy conservation means that the total energy (electron + ion) is constant in time:
\begin{align}
\label{eq:ei_energy}
\frac{3}{2}N_i\frac{\partial T_i} {\partial t} + \int_0^\infty \varepsilon \frac{\partial J_{ei}} {\partial \varepsilon } \, d\varepsilon = 0.
\end{align}
Hence, to impose energy conservation, one can update the ion energy (or temperature) according to the discrete form of the integral in Eq.~(\ref{eq:ei_energy}). Energy conservation for the ee collision operator is expressed as:
\begin{align}
\int_0^\infty \varepsilon \frac{\partial J_{ee}} {\partial \varepsilon } \, d\varepsilon = 0.
\end{align}
It is less straightforward to satisfy this condition at the discrete level, even for the classical FP equation. We address this issue in the next section.

\section{Numerical Method}
\subsection{Energy Discretization}
In this section, we describe a DG method for solving the qFP equation~(\ref{eq:qfp}). Let us discretize the solution domain into $N$ non-overlapping energy groups, $\varepsilon \in \left[ \varepsilon_{i-1/2}, \varepsilon_{i+1/2} \right] = \varOmega_i$, where $i \pm 1/2$ denotes group boundaries. In each energy group, we express the local solution of $f$ as an expansion in terms of an orthonormal set of basis functions $U_p (\varepsilon) $:
\begin{align}
\label{eq:expansion}
\varepsilon \in \varOmega_i: f_i (\varepsilon,t) = \sum_{p=0}^{p_{\max}} \hat{f}_{i,p} (t) U_p (\varepsilon).
\end{align}
It is convenient to introduce a mapping from $\varOmega_i$ to $[-1,1]$ using the transformation $x = \frac{2(\varepsilon - \varepsilon_i)}{\varDelta_i}$, where $\varepsilon_i$ denotes the energy at the center of group $i$ and $\varDelta_i$ the width of the group. Hereafter to simply the expressions, we shall occasionally interchange the variables $x$ and $\varepsilon$. In this work we choose the normalized Legendre polynomials for the basis functions, i.e., $U_p (x) = \sqrt{p+\frac{1}{2}} P_p (x)$ where $P_p (x)$ is the regular Legendre polynomial. The orthogonality relationship for the basis functions is:
\begin{align}
\label{eq:orthogonality}
\int_{-1}^1 U_p (x) U_q(x) dx = \delta_{pq}.
\end{align}
{From (\ref{eq:expansion}) and (\ref{eq:orthogonality}), we can show that any discrete $p$-th order moment of the distribution function can be evaluated exactly as linear combination of expansion coefficients $\hat{f}_q$ ($q \leq p$).} For example, it is straightforward to show that number density and energy of a group are linear combinations of only the first two coefficients:
\begin{subequations}
\label{eq:ni_ei}
\begin{align}
\label{eq:ni}
\bar{n}_i &= \int_{\varOmega_i} f \, d\varepsilon = \frac{\varDelta_i}{\sqrt{2}} \hat{f}_{i,0}, \\
\bar{e}_i &= \int_{\varOmega_i} f \varepsilon \, d\varepsilon = \frac{\varDelta_i \varepsilon_i}{\sqrt{2}} \hat{f}_{i,0} + \frac{\varDelta_i^2}{2\sqrt{6}} \hat{f}_{i,1},
\end{align}
\end{subequations}
where $\int_{\varOmega_i}$ denotes an integral over group $i$. Eq.~(\ref{eq:ni_ei}) indicates that the time evolution of number density and energy of the system can be correctly described if the expansion in Eq.~(\ref{eq:expansion}) is carried to at least second order ($p_{\max}=1$). {This means that density and energy of each group, or more specifically, some linear combination of them, are included as part of the solution vector.} As will be shown later, this additional degree of freedom also gives us the ability to enforce discrete energy conservation. Although the DG method presented in this section can be generalized to expansion of arbitrary order, the numerical results presented in this work is only for $p_{\max} = 1$. {Although a higher-order expansion is more accurate, it is not necessarily advantageous since the computational requirements increase with the maximum order of the basis functions. In addition, the higher-order moments lose physical meaning; for example, the second-order moment could be associated with energy fluctuations about the mean, but there is no conservation law associated with these higher-order quantities. This makes coupling with other physics, e.g., inelastic processes, more complicated.} In the past, we demonstrated that this method can also be applied to model inelastic collisions by a Boltzmann collision operator \cite{le_conservative_2017}. The extension of the Boltzmann collision operator to take into account degeneracy is left for future work.

For the sake of generality, we now describe the DG discretization scheme using a generic flux function $J$. The difference in the treatment of ei and ee collisions will be pointed out wherever appropriate. The time rate of change of the expansion coefficients is obtained by multiplying Eq.~(\ref{eq:qfp}) by the basis function and integrating over group $i$:
\begin{align}
\label{eq:dg}
\frac{\varDelta_i}{2} \frac{d \hat{f}_{i,p}}{dt} &= \left[ \hat{J}_{i+1/2} U_p(1) - \hat{J}_{i-1/2} U_p(-1) \right] - \int_{\varOmega_i} J \frac{dU_p}{d\varepsilon} d\varepsilon.
\end{align}
Here $\hat{J}$ is the numerical flux at an interface, which can be decomposed into a convective and diffusive parts:
\begin{align}
\hat{J}_{i+1/2} = \hat{J}^C_{i+1/2} + \hat{J}^D_{i+1/2}.
\end{align}
The convective and diffusive fluxes for ei collision term are:
\begin{subequations}
\label{eq:jei}
\begin{align}
\left[ \hat{J}^C_{i+1/2} \right]_{ei} &= \gamma_{ei} \varepsilon_{i+1/2}^{-1/2} f^C_{i+1/2} \left( 1- \tilde{f}^C_{i+1/2} \right) , \\
\left[ \hat{J}^D_{i+1/2} \right]_{ei} &= \gamma_{ei} T_i  \frac{\partial}{\partial \varepsilon} \left( \frac{f^D}{\sqrt{\varepsilon}} \right)_{i+1/2}.
\end{align}
\end{subequations}
The convective and diffusive fluxes for ee collision term are:
\begin{subequations}
\label{eq:jee}
\begin{align}
\left[ \hat{J}^C_{i+1/2} \right]_{ee} &= \gamma L_{i+1/2} \varepsilon_{i+1/2}^{-1/2} f^C_{i+1/2} \left( 1- \tilde{f}^C_{i+1/2} \right) , \\
\left[ \hat{J}^D_{i+1/2} \right]_{ee} &= \gamma K_{i+1/2} \frac{\partial}{\partial \varepsilon} \left( \frac{f^D}{\sqrt{\varepsilon}} \right)_{i+1/2}.
\end{align}
\end{subequations}
In Eqs.~(\ref{eq:jei}) and (\ref{eq:jee}), we introduce average values of the distribution function at an interface for convective and diffusive fluxes, denoted by the superscript $C$ and $D$ respectively. For convective fluxes, the average value at the interface is defined using the Chang-Cooper method \cite{chang_practical_1970}:
\begin{align}
\label{eq:CC}
f^C_{i+1/2} = 	\delta_{i+1/2} f^-_{i+1/2} + \left( 1- \delta_{i+1/2} \right) f^+_{i+1/2},
\end{align}
where $f^-_{i+1/2}$ denotes the value of $f$ evaluated at the right boundary of cell $i$ and $f^+_{i+1/2}$ the value of $f$ evaluated at the left boundary of cell $i+1$. The weighting coefficient $\delta_{i+1/2}$ is defined as:
\begin{align}
\delta_{i+1/2} = \frac{1}{\omega_{i+1/2}} - \frac{1}{e^{\omega_{i+1/2}} - 1},
\end{align}
where $\omega_{i+1/2}$ is related to the ratio of the convective to the diffusive coefficient. Note that $f^C$ is defined differently for ei and ee collisions. Here we use a version of Chang-Cooper average for energy grid similar to Buet and Le Thanh \cite{buet_eficient_2008}. For ei collision, $\omega_{i+1/2} = \frac{\varepsilon_{i+1} - \varepsilon_{i}}{T_i}$ and for ee collision, $\omega_{i+1/2} = \frac{(\varepsilon_{i+1} - \varepsilon_{i}) L_{i+1/2} }{K_{i+1/2}}$. We point out that the blocking factor is omitted from these definitions, because it directly depend on the solution at the interface. This simplification does not introduce significant error when the distribution is weakly degenerate. For a degenerate distribution, we would expect some errors in the energy range $\varepsilon \lesssim \mu + T_e$. However, a Chang-Copper type of average in this range is not justifiable so any averaging of the form (\ref{eq:CC}) where $\delta_{i+1/2} \in [0,1]$ is reasonable. {We emphasize that this simplification is only applied to the calculation of the weighting factors $\omega_{i+1/2}$ and $\delta_{i+1/2}$.}

{We note that a standard DG scheme does not guarantee positive distribution inside all groups when $p_{\max} \ge 1$. This is a known issue for the use of DG schemes to solve kinetic equations, where the particle distribution always remains a non-negative scalar. Application of a positivity preserving limiter \cite{zhang_maximum_2011} can alleviate this problem but for the case of FP equation it will destroy conservation properties. The adaptation of the Chang-Cooper flux, motivated by its use in standard FV schemes for FP equation, is a way to mitigate this problem, by taking into account the exponential fall off of the distribution to avoid overestimating the flux.}

The average value of $f$ for the diffusive flux is calculated according to the recovery-based DG scheme of van Leer et al. \cite{van_leer_discontinuous_2007}. The procedure is briefly summarized here. For each interface $i+1/2$, we first recover a polynomial $g \equiv g (f_i,f_{i+1})$ that is continuous across two adjacent cells $i$ and $i+1$, {i.e., over $\varOmega_i \cup \varOmega_{i+1}$,} from a $L_2$ minimization:
\begin{subequations}
\begin{align}
\int_{\varOmega_i} (g - f_i) U_p \, d\varepsilon&= 0, \\
\int_{\varOmega_{i+1}} (g - f_{i+1}) U_p \, d\varepsilon&= 0.
\end{align}
\end{subequations}
If $f$ is represented by a polynomial of order $p_{\max}$, the recovered polynomial $g$ is of order $2p_{\max} + 1$. The boundary value and derivatives can then be computed from the recovered polynomial.

The integral over energy group in Eq.~(\ref{eq:dg}) can be approximated by Gaussian quadrature:
\begin{align}
\label{eq:quad}
\int_{\varOmega_i} J \frac{dU_p}{d\varepsilon} d\varepsilon \simeq \sum_n w_n J \left( x_n \right) \frac{dU_p}{dx},
\end{align}
where $n$ denotes a quadrature point and $w_n$ its weight. The approximation in Eq.~(\ref{eq:quad}) is used for ei collisions. For ee collisions, it is more advantageous to rewrite Eq.~(\ref{eq:quad}) for $p=1$ into a flux form to get exact energy conservation. This will be shown in the next section.

{
We make a remark regarding the order of accuracy of the proposed scheme. For linear problems, the order of accuracy of the convective term is $p_{\max}+1$, while the order of accuracy of the diffusive term, using the recovery-based approach, is $2^{p_{\max}+1}$ \cite{van_leer_discontinuous_2007}. This exponential accuracy is only obtained in one dimension which is the case here. For higher dimensions, the order of accuracy for recovery-based DG schemes is $2(p_{\max}+1)$. Therefore the overall combined accuracy should be dictated by the convective term. With $p_{\max}=1$, we can expect overall second order accuracy for the solution of the qFP equation. This will be checked numerically in Sec. \ref{sec:thermalization}.
}

\subsection{Conservation Properties}
It is important to ensure that the numerical method respect any conservation law associated with the collision operators. It is easy to show that the present numerical scheme is density conserving. Summing Eq.~(\ref{eq:dg}) for $p=0$ over all groups and using Eq.~(\ref{eq:ni}), the time rate of change of the total electron density is:
\begin{align}
\label{eq:density}
\frac{dN_e}{dt} = \sum_{i=1}^{N} \frac{d \bar{n}_i}{dt} = \hat{J}_{N+1/2} - \hat{J}_{1/2} .
\end{align}
Since the imposed boundary conditions are that particle fluxes vanish at domain boundaries, the sum in Eq.~(\ref{eq:density}) is exactly zero. This is true for both ei and ee collision operators. For ee collisions, we can show that total electron energy is conserved in the continuous limit. At the discrete level, the time evolution of the total energy can be written as:
\begin{align}
\label{eq:energy}
\left[ \frac{dE_e}{dt} \right]_{ee} = \sum_{i=1}^{N} \frac{d \bar{e}_i}{dt} = \hat{J}_{ee,N+1/2} \varepsilon_{N+1/2} - \hat{J}_{ee,1/2} \varepsilon_{1/2} -  \sum_i \int_{\varOmega_i} J_{ee} \, d\varepsilon.
\end{align}
Again the flux terms are zeros due to the imposed boundary conditions. To conserve total energy, it is essential to ensure that the discrete sum in Eq.~(\ref{eq:energy}) is exactly zero. Note that the integral in Eq.~(\ref{eq:energy}) shows up on the right hand side of Eq.~(\ref{eq:dg}) for $p=1$ (second term). This integral cannot be calculated exactly, and numerical approximation using Gaussian quadrature always introduce some errors. To achieve exact energy conservation, we will rewrite this term into a flux form following a similar approach highlighted in \cite{le_conservative_2017} for the classical Fokker-Planck equation. Substituting the expression for $J_{ee}$ into the integral in Eq.~(\ref{eq:energy}) and integrating by part, we obtain:
\begin{align}
\label{eq:flux}
\int_{\varOmega_i} J_{ee}  d\varepsilon = \gamma \left[ K \varepsilon^{-1/2} f \right]^{i+1/2}_{i-1/2} - \gamma \int_{\varOmega_i}  \left[ \frac{\partial K}{\partial \varepsilon}  - L (1- \tilde{f}) \right] \varepsilon^{-1/2} f \, d\varepsilon.
\end{align}
{Let us examine the second term on the right hand side of Eq.~(\ref{eq:flux}). We first differentiate Eq.~(\ref{eq:ee_k}) with respect to $\varepsilon$ to obtain an expression for $\frac{\partial K}{\partial \varepsilon}$. The derivative of the integrals can be evaluated using the first fundamental theorem of calculus. After some simplification, we arrive at:
\begin{align}
\label{eq:ee_dkde}
\frac{\partial K}{\partial \varepsilon} 
&= 3\varepsilon^{1/2} \int_\varepsilon^\infty f(1-\tilde{f}) \varepsilon'^{-1/2} \, d \varepsilon'.
\end{align}
To simplify the notation, let us define $h(\varepsilon) \equiv 1-\tilde{f} (\varepsilon)$. Using (\ref{eq:ee_dkde}) and (\ref{eq:ee_l}), the second term on the right hand side of Eq.~(\ref{eq:flux}) becomes:
\begin{align}
\label{eq:flux1b}
\int_{\varOmega_i} J_{ee}  d\varepsilon &\varpropto - \gamma \int_{\varOmega_i}  \left[ 3 f \int_\varepsilon^\infty f h \ \varepsilon'^{-1/2} \, d \varepsilon'  - 3 f h \ \varepsilon^{-1/2} \int_0^{\varepsilon} f \, d \varepsilon' \right]  \, d\varepsilon \nonumber \\
& = - \gamma \int_{\varOmega_i}  \left[ 3 f \int_\varepsilon^\infty f h \ \varepsilon'^{-1/2} \, d \varepsilon'  + \left( \frac{\partial}{\partial \varepsilon} \int_{\varepsilon}^\infty 3  f h \varepsilon^{-1/2} \right) \left( \int_0^{\varepsilon} f \, d \varepsilon' \right) \right]  \, d\varepsilon.
\end{align}
The second term on the right hand side of (\ref{eq:flux1b}) can be integrated by part. This results in two terms, one of which cancels the first term on the right hand side of (\ref{eq:flux1b}) exactly. Hence, we arrive at:
\begin{align}
\label{eq:flux1c}
\int_{\varOmega_i} J_{ee}  d\varepsilon \varpropto - \gamma \left[ \left( \int_{\varepsilon}^{\infty} 3 \varepsilon^{'1/2} f h \, d\varepsilon' \right) \left( \int_{0}^\varepsilon f \, d\varepsilon' \right) \right]^{i+1/2}_{i-1/2}.
\end{align}
Putting this result back to Eq.~(\ref{eq:flux}), we obtain:
\begin{align}
\label{eq:flux2}
\int_{\varOmega_i} J_{ee} \, d\varepsilon = \gamma \left[ K \varepsilon^{-1/2} f \right]^{i+1/2}_{i-1/2} - \gamma \left[ \left( \int_{\varepsilon}^{\infty} 3 \varepsilon^{'1/2} f \left( 1- \tilde{f} \right) \, d\varepsilon' \right) \left( \int_{0}^\varepsilon f \, d\varepsilon' \right) \right]^{i+1/2}_{i-1/2}.
\end{align}
We can see that the rate of change in total energy expressed in Eq.~(\ref{eq:energy}) with the integral given by Eq.~(\ref{eq:flux2}) is not exactly zero. When summing over all groups, there is a single term remained in the expression:
\begin{align}
\label{eq:energy2}
\sum_{i=1}^{N} \frac{d \bar{e}_i}{dt} = -\gamma K_{N+1/2} \varepsilon_{N+1/2}^{-1/2} f_{N+1/2}
\end{align}
Eq.~(\ref{eq:energy2}) gives the error in total energy at the discrete level.} This is a direct consequence of the finite truncation of the solution domain. In order to exactly conserve total energy, we propose the following approximate expression for the flux integral:
\begin{align}
\label{eq:eflux}
\int_{\varOmega_i} J_{ee} \, d\varepsilon \simeq \gamma \left[ K \varepsilon^{-1/2} \left( f - f_{N+1/2} \right) \right]^{i+1/2}_{i-1/2} - \gamma \left[ \left( \int_{\varepsilon}^{\infty} 3\varepsilon^{'1/2} f \left( 1- \tilde{f} \right) \, d\varepsilon' \right) \left( \int_{0}^\varepsilon f \, d\varepsilon' \right) \right]^{i+1/2}_{i-1/2},
\end{align}
where $f_{N+1/2}$ is the solution at the last interface. This additional term acts as a correction to the energy flux to ensure conservation. Another acceptable approximation is:
\begin{align}
\label{eq:eflux2}
\int_{\varOmega_i} J_{ee} \, d\varepsilon \simeq \gamma \left[ K \left( \varepsilon^{-1/2} f - \varepsilon^{-1/2}_{N+1/2} f_{N+1/2} \right) \right]^{i+1/2}_{i-1/2} - \gamma \left[ \left( \int_{\varepsilon}^{\infty} 3\varepsilon^{'1/2} f \left( 1- \tilde{f} \right) \, d\varepsilon' \right) \left( \int_{0}^\varepsilon f \, d\varepsilon' \right) \right]^{i+1/2}_{i-1/2}.
\end{align}
It is straightforward to verify that the sum of Eq.~(\ref{eq:eflux}) or (\ref{eq:eflux2}) over all energy groups is exactly zero. Hence, total energy is exactly conserved.

\subsection{Time Integration}
The semi-discrete system (\ref{eq:dg}) is integrated by a backward Euler method. It can be put into the following form:
\begin{align}
\label{eq:be}
\frac{\hat{f}_{i,p}^{n+1} -\hat{f}_{i,p}^{n} }{\varDelta t} = R_{i,p} (\hat{f}^{n+1}),
\end{align}
where $R_{i,p}$ contains the right hand side of Eq.~(\ref{eq:dg}) evaluated at the next time step. Note that the factor $\varDelta_i/2$ on the left hand side of Eq.~(\ref{eq:dg}) is absorbed into $R_{i,p}$. Linearizing Eq.~(\ref{eq:be}) about the solution at the current time step leads to:
\begin{align}
\label{eq:linearization}
R_{i,p} (\hat{f}^{n+1}) \simeq R_{i,p} (\hat{f}^{n}) + \sum_{j,q} \frac{\partial R_{i,p} }{\partial f_{j,q}} \left( \hat{f}_{j,q}^{n+1} -\hat{f}_{j,q}^{n} \right),
\end{align}
where $\sum_{j,q}$ denotes a sum over all group and polynomial indices. Since the convective and diffusive coefficients for ee collisions involve integrals of the distribution function over all energy groups, the Jacobian $\frac{\partial R_{i,p} }{\partial f_{j,q}}$ is generally a dense matrix. {\ref{app:kl} gives detailed analytical expressions for the these coefficients and their derivatives with respect to the solution vector, which are used to construct the Jacobian. We also describe in \ref{app:kl} an efficient method to evaluate these coefficients and derivatives during the computation.} 

Substituting Eq.~(\ref{eq:linearization}) into (\ref{eq:be}), we obtain a linear system to be solved at each time step:
\begin{align}
\label{eq:be2}
\sum_{j,q} \left( \varDelta t^{-1} \delta_{ij} \delta_{pq} - \frac{\partial R_{i,p} }{\partial f_{j,q}}\right) \left( \hat{f}_{j,q}^{n+1} -\hat{f}_{j,q}^{n} \right) = R_{i,p} (\hat{f}^{n}) ,
\end{align}
where $\delta_{ij}$ and $\delta_{pq}$ are Kronecker deltas. {We note that inverting Eq.~(\ref{eq:be2}) constitutes the most computationally intensive part of the calculation since this operation scales as $O(N_{\text{dof}}^3)$ where $N_{\text{dof}}=N(p_{\max}+1)$. Using the method in \ref{app:kl}, the computational cost associated with evaluating the right hand side of (\ref{eq:be2}) and the Jacobian is very small}. We note that although energy conservation is exact in Eq.~(\ref{eq:be}), the linearization in Eq.~(\ref{eq:linearization}) can introduce some numerical errors. Iterative improvement can be done to converge any non-linearity. However, numerical tests show that the linearization errors are very small and cannot be distinguished from numerical round-off errors. The results shown in this paper are obtained from solving the linearized system (\ref{eq:be2}). 

{We note that the backward Euler method is first order accurate in time. Since the method is fully implicit, we can relax the stability constraint and only impose time step restriction based on accuracy requirement. The time step size is estimated by limiting maximum rate of change of density within an energy group using the following expression:
\begin{equation}
\Delta t^{\text{new}} = \min_{\substack{ 1\le i \le N \\ \bar{n}^{n+1}_i > \beta N_e}} \frac{\alpha  \Delta t }{|1 - \bar{n}^n_i/\bar{n}^{n+1}_i|},
\label{eq:timestep}
\end{equation}
where $\Delta t^{\text{new}}$ is an estimated time step for the next cycle. The $\min$ operation is only performed over groups that have non-negligible densities. Here we set $\alpha = 0.1$ and $\beta = 10^{-8}$. For all simulations shown in this work, the time step is estimated using Eq.~(\ref{eq:timestep}). The only time we use constant time step size is at the beginning of Sec. \ref{sec:thermalization} to check numerical convergence.}

\section{Simulation Results}
{
\subsection{Electron Thermalization}
\label{sec:thermalization}
In this section, we present several numerical tests to validate the numerical method. We first consider the thermalization of a monoenergetic distribution to demonstrate the order of accuracy of the scheme. Here only the ee collision term is examined. Since both ee and ei collision terms are discretized the same way, we expect similar order of accuracy. The electron distribution in unit cm$^{-3}$-eV$^{-1}$ is initialized as follows:
\begin{align}
f(\varepsilon,t=0) &= \frac{1}{\sqrt{\pi}} \frac{N_e}{\delta_h}  \exp \left[ - \frac{ \left( \varepsilon - \varepsilon_h \right)^2 }{\delta_h^2} \right]
\end{align}
where $N_e = 10^{21}$ cm$^{-3}$, $\varepsilon_h=0.5$ eV and $\delta_h = 0.12$ eV. This problem is similar to the one from Epperlein\cite{epperlein_implicit_1994}, but here the electrons are degenerate due to high density and low temperature. At this density, the Fermi energy $E_F = 0.34$ eV is comparable to the total energy $E_e = 0.5$ eV. Let us can define a reference time using the ee mean collision time:
\begin{align}
\label{eq:tauee}
\tau_{ee}^{-1} = \frac{4}{3} \frac{e^4 m}{\pi \hbar^3} \frac{1}{1+e^{-\mu^\star /T^\star}_e} \ln \varLambda_{ee},
\end{align}
where $\mu^\star/T^\star_e = 0.5$ and $T^\star_e = 0.28$ eV are the equilibrium values calculated from the initial distribution. In the classical limit, we recover the standard expression for $\tau_{ee}$ \cite{huba_nrl_2016}.

We use an energy grid from 0 to 3 eV with constant energy spacings. Fig. \ref{fig:relax_epperlein} shows the typical evolution of the occupation number $\tilde{f} \varpropto f/\sqrt{\varepsilon}$ at different times as the distribution relaxes toward a Fermi-Dirac distribution. The classical Maxwellian distribution is also shown to highlight the difference in the equilibrium spectrum when degeneracy is taken into account. The numerical results in Fig. \ref{fig:relax_epperlein} are obtained using 256 groups and a constant time step size of $\Delta t = 0.01 \tau_{ee}$. To check the order of accuracy of the scheme, Figs. \ref{fig:econv} and \ref{fig:tconv} show the $L_p$ ($p=1,2,\infty$) norms of the error, defined as the difference between the numerical and a reference solution of the distribution. Fig.~\ref{fig:econv} shows the error as a function of the number of energy groups $N$ using a constant $\Delta t = 0.01 \tau_{ee}$. The reference solution is calculated at $t=0.1 \tau_{ee}$ using $N=1024$ groups. Fig.~\ref{fig:tconv} shows the error as a function of time step size for a constant number of groups $N=128$. The reference solution is calculated at $t=0.05 \tau_{ee}$ using $\Delta t = 10^{-4} \tau_{ee}$. The results from Figs. \ref{fig:econv} and \ref{fig:tconv} clearly indicate second order convergence in energy group size and first order temporal convergence.

\begin{figure}
\centering
\includegraphics[scale=1]{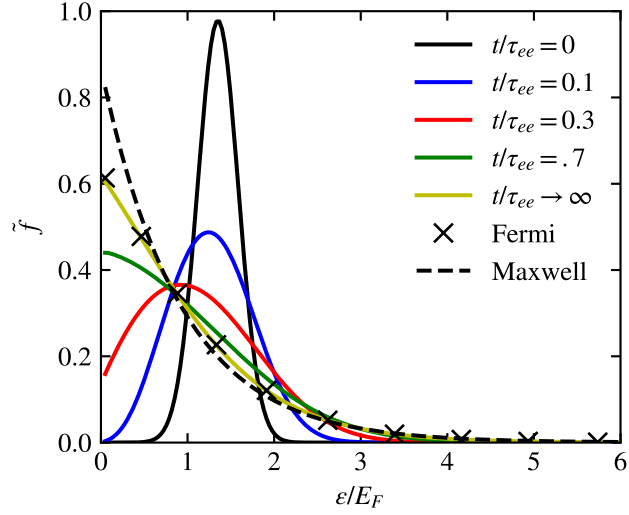}
\caption{Relaxation of a monoenergetic electron energy distribution functions at different times indicated by different colors. The symbols represent the expected equilibrium Fermi-Dirac distribution while the dashed line is the classical equilibrium Maxwellian distribution}
\label{fig:relax_epperlein}
\end{figure}

\begin{figure}
\begin{subfigure}[t]{0.5\textwidth}
\includegraphics[scale=.8]{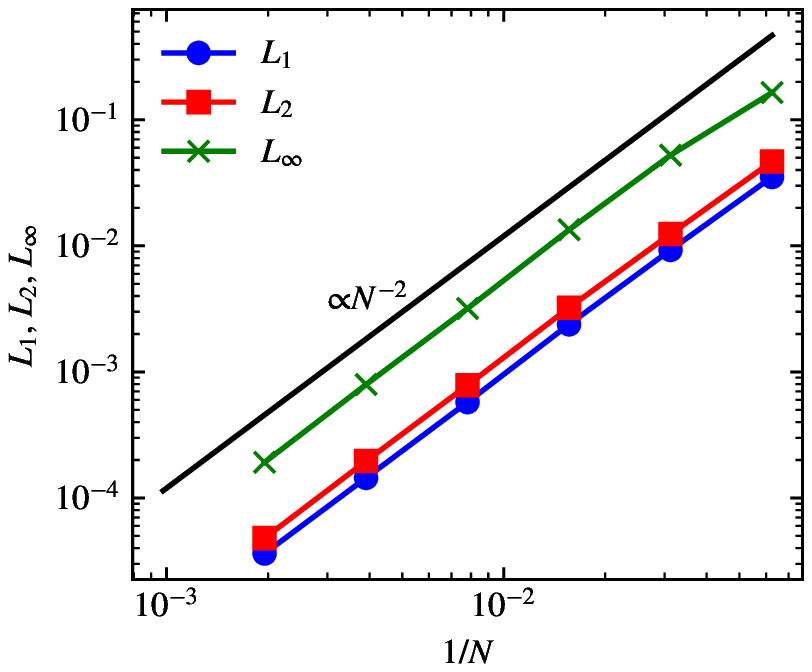}
\caption{}
\label{fig:econv}
\end{subfigure}
~
\begin{subfigure}[t]{0.5\textwidth}
\includegraphics[scale=.8]{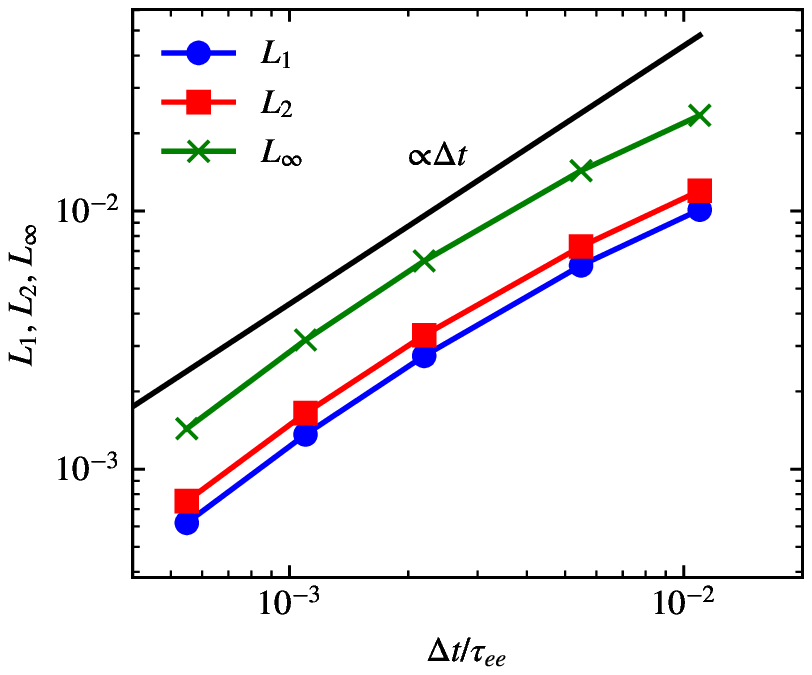}
\caption{}
\label{fig:tconv}
\end{subfigure}
\caption{The $L_1$, $L_2$ and $L_\infty$ errors of the distribution function for the problem shown in Fig. \ref{fig:relax_epperlein} as function of (a) number of groups $N$ and (b) time step size $\Delta t$.}
\label{fig:conv}
\end{figure}
}

The next test case is a more challenging problem requiring a nonuniform energy grid for efficiency. This problem is a simple model describing thermalization of excited electrons in metals produced by x-ray absorption \cite{kitamura_thermalization_2019}. The relaxation of these high energy electrons due to interaction with the conduction electrons are modeled using the qFP equation. Here the conduction electrons are treated as a uniform electron gas at zero temperature. The electron distribution in unit cm$^{-3}$-eV$^{-1}$ is initialized as:
\begin{subequations}
\begin{align}
f(\varepsilon,t=0) &= 4\pi \left( \frac{2m}{h^2} \right)^{3/2} \sqrt{\varepsilon} \tilde{f}_0 (\varepsilon), \\
\tilde{f}_0 (\varepsilon) &=
\begin{cases}
      1 & \text{if $\varepsilon \le E_F$}\\
      \frac{2 E_F^{3/2} x_h}{3\sqrt{\pi} \delta_h} \frac{1}{\sqrt{\varepsilon}}\exp \left[ - \frac{ \left( \varepsilon - \varepsilon_h \right)^2 }{\delta_h^2} \right] & \text{if $\varepsilon>E_F$}
    \end{cases},      
\end{align}
\end{subequations}
where $E_F=7$ eV, $x_h = 1.6 \times 10^{-3}$, $\varepsilon_h = 1$ keV and $\delta_h = 10$ eV. The initial distribution consists of a population of cold dense electrons at zero temperature and a small fraction of high energy electrons with a total electron density of $8.47 \times 10^{22}$ cm$^{-3}$. The high energy electrons are initialized as a Gaussian distribution centered at 1 keV. The ee mean collision time is defined using Eq.~(\ref{eq:tauee}) with $\mu^\star/T^\star_e = 2.66$ and $T^\star_e = 2.36$ eV. In this test, ei collision is turned off. 

{In this simulation, we use an energy grid with nonuniform spacing and clustered points near the Gaussian to ensure that the particle distributions at low and high energies are adequately resolved. The main strategy is to divide the physical domain into a low and high energy portion. For each portion, we utilize the formula describe in \ref{app:grid} to generate a grid that has grid spacings successively increased by a constant ratio. The energy grid for this problem is defined as:
\begin{align}
\label{eq:grid1}
\varepsilon_{i+1/2} & = 
\begin{cases}
0, & \text{for}\ i=0 \\
\varepsilon_{i-1/2} + r_1^{i-1} \Delta \varepsilon_1, & \text{for} \ i=1,2,\cdots, M\\
\varepsilon_{i-1/2} + r_2^{i-1} \Delta \varepsilon_2, & \text{for} \ i=M+1,M+2,\cdots, N\\
\end{cases}
\end{align}
We ran the simulation with successively refined grids (64, 128 and 256 energy groups). Table \ref{tab:sim1} lists values of $M$, $r_1$ and $r_2$ for different values of $N$.

\begin{table}[ht]
\centering
\begin{tabular}{|c|c|c|c|}
\hline 
$N$ & $M$ & $r_1$ & $r_2$ \\ 
\hline 
$64$ & $52$ & $1.18$ & $1.10$ \\ 

$128$ & $98$ & $1.05$ & $1.05$ \\ 

$256$ & $185$ & $1.02$ & $1.01$ \\ 
\hline 
\end{tabular}
\caption{Grid generation parameters for simulations shown in Figs.~\ref{fig:relax_ee_edist} - \ref{fig:relax_ee_error}}
\label{tab:sim1}
\end{table}
In Eq.~(\ref{eq:grid1}), $\Delta \varepsilon_1$ is calculated from Eq.~(\ref{eq:egrid}) using $N_\varepsilon=M$, $\varepsilon_{\min}=10^{-2}$ eV and $\varepsilon_{\max}=900$ eV, and $\Delta \varepsilon_2$ is calculated using $N_\varepsilon=N-M$, $\varepsilon_{\min}=900$ eV and $\varepsilon_{\max}=2000$ eV. We obtained grid-independent results with 256 groups.

}

\begin{figure}
\begin{subfigure}[t]{0.5\textwidth}
\includegraphics[scale=.8]{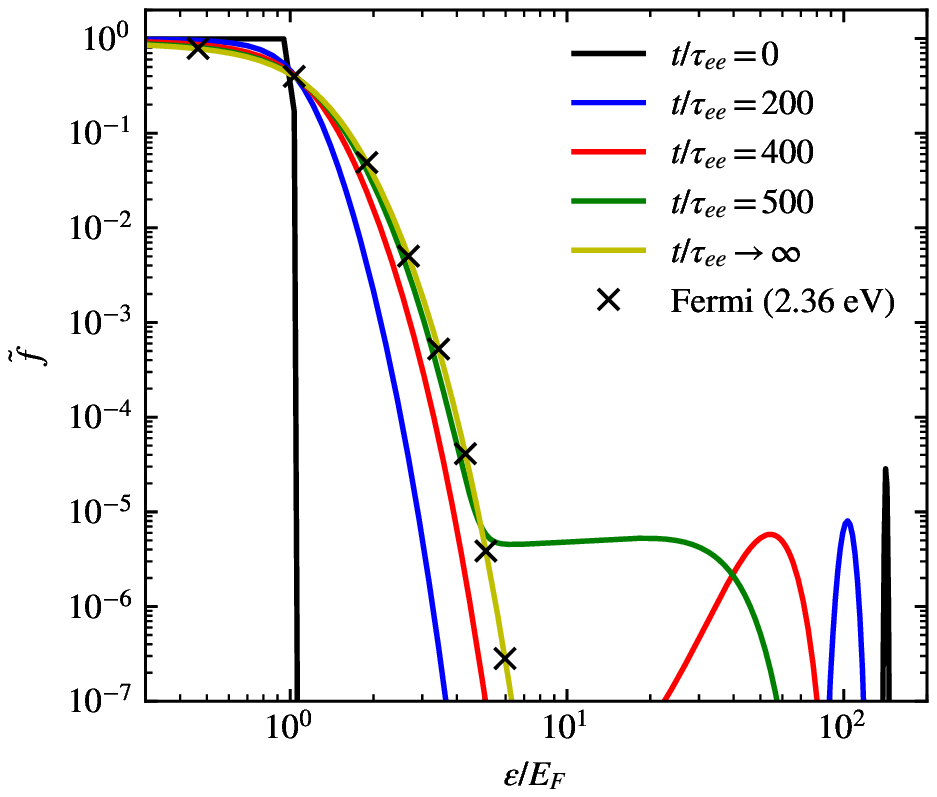}
\caption{}
\label{fig:relax_ee_edista}
\end{subfigure}
~
\begin{subfigure}[t]{0.5\textwidth}
\includegraphics[scale=.8]{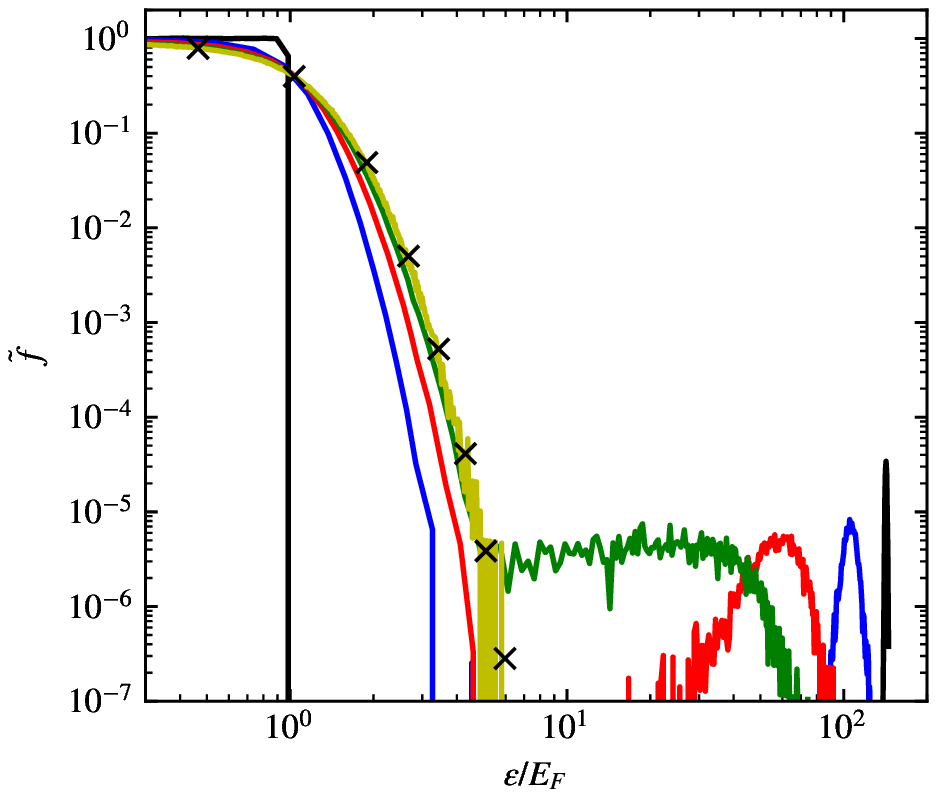}
\caption{}
\label{fig:relax_ee_edistb}
\end{subfigure}
\caption{Thermalization of high energy electrons in a uniform electron gas at zero temperature. $N_e=8.47 \times 10^{22}$ cm$^{-3}$, $E_F=7$. Time evolution of the EEDF using (a) qFP equation and (b) Monte Carlo collision method. Different colors indicate different times, and the symbols represent an exact Fermi-Dirac distribution at $\mu/T_e = 2.66$ and $T_e = 2.36$ eV. {The qFP simulation uses 256 groups, and the Monte Carlo simulation uses 2 million particles.}}
\label{fig:relax_ee_edist}
\end{figure}

\begin{figure}
\centering
\includegraphics[scale=.8]{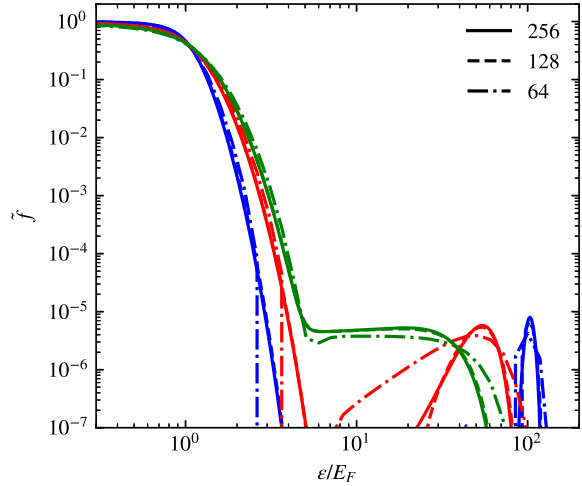}
\caption{Electron energy distribution functions at $t/\tau_{ee} =$ 200, 400 and 500 using three different energy grids (64, 128 and 256 groups). Different colors indicates different times similarly to Fig.~\ref{fig:relax_ee_edist}. Different linestyles indicate results from different grids. {The solid and dashed lines are almost indistinguishable}.}
\label{fig:relax_ee_edist_grid}
\end{figure}

To validate the qFP results, we perform the same simulation but using a Monte Carlo collision (MCC) method to evolve the EEDF instead. The MCC algorithm is based on a modified Takizuka-Abe method introduced by Turrell et al. \cite{turrell_monte_2013}. Due to the large dynamic range both in density and energy, the Monte Carlo simulation uses a large number of particles (2 million). The comparison of the EEDF's { obtained with 256 groups} at several times are shown in Fig.~\ref{fig:relax_ee_edist}. The EEDF's obtained from the qFP equation agree well with those from the Monte Carlo simulation. In particular, the relaxation of the high-energy electrons between two approaches are identical. However, it is clear that the Monte Carlo solutions suffer from numerical noises, and the features in the high energy tail of the EEDF are better resolved with the qFP approach. At steady-state, the EEDF's from both methods approach the correct equilibrium limit. To demonstrate that the numerical results are grid-independent, Fig.~\ref{fig:relax_ee_edist_grid} shows the distributions at different times using three different of grids (64, 128 and 256 groups). It is evident that the results obtained with 128 groups are already very close to the converged solution. Fig.~\ref{fig:relax_ee_hfunct} shows that the relative $H$ function, calculated from the distribution, is always decreasing and approaching its minimum at equilibrium. Here we define the relative $H$ function similar to \cite{hu_numerical_2012}:
\begin{align}
\label{eq:h}
H = N_e^{-1} \int_0^\infty f \ln \left[ \frac{\tilde{f}}{1-\tilde{f}} \frac{1-\tilde{f}^\star}{\tilde{f}^\star} \right] \, d\varepsilon,
\end{align}
This result is consistent with H-theorem \cite{hu_numerical_2012,daligault_quantum_2016}. Fig.~{\ref{fig:relax_ee_error} shows the normalized energy errors both globally (solid) and per time step (dashed). The errors are extremely small ($< 10^{-14}$), confirming that the numerical scheme is energy conserving. This also illustrates the point that the linearized system (\ref{eq:be2}) is a robust approximation to (\ref{eq:be}).

\begin{figure}
\centering
\includegraphics[scale=1]{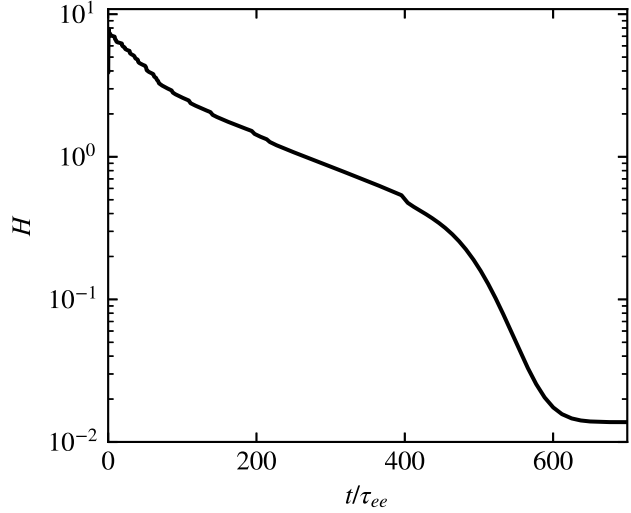}
\caption{Evolution of the relative $H$ function.}
\label{fig:relax_ee_hfunct}
\end{figure}

\begin{figure}
\centering
\includegraphics[scale=1]{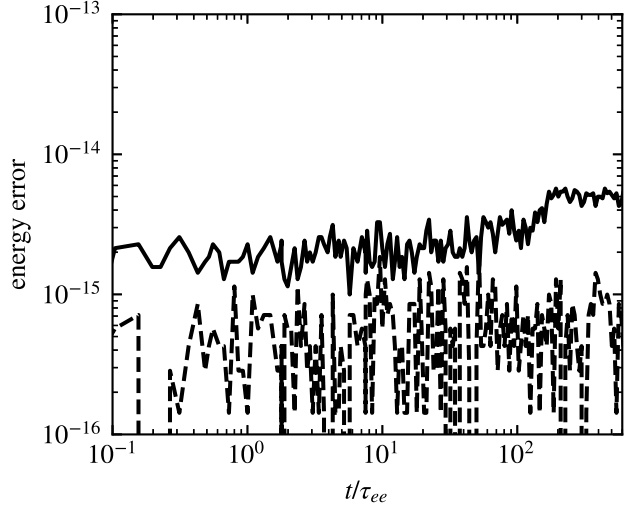}
\caption{Total energy error (solid) and error per time step (dash).}
\label{fig:relax_ee_error}
\end{figure}

\subsection{Electron-ion Equilibration}
The next two test cases model electron-ion equilibration process in a dense hydrogen plasma ($Z=1$) using both ei and ee collision operators. For simplicity, we assume constant ion temperature, so we expect that at steady state the EEDF becomes a Fermi-Dirac distribution at this temperature. The first test is a heating problem where the distribution is evolved from a degenerate to a classical distribution. The second test is the opposite where the EEDF goes from a classical to a degenerate distribution. {The energy grid are generated according to the formula:
\begin{align}
\varepsilon_{i+1/2} & = 
\begin{cases}
0, & \text{for}\ i=0 \\
\left( \frac{\varepsilon_{\max}}{ \varepsilon_{\min}} \right)^{\frac{1}{N}} \varepsilon_{i-1/2}, & \text{for} \ i=1,2,\cdots,N
\end{cases}
\end{align}
where $N$ is the number of groups, $\varepsilon_{\min}=10^{-2}$ eV and $\varepsilon_{\max}= 10^3$ eV. Both simulations use the same energy grid consisting of 64 groups.}

{In the limit where the ee collision time is much faster than ei equilibration time, the EEDF can be approximated as a Fermi-Dirac distribution at all times (instantaneous thermalization) and the ei energy transfer rate can be obtained from direct evaluation of the energy moment of the full qFP collision integral \cite{daligault_quantum_2016}. In many situations, we can assume that the ion thermal velocity is much smaller than the electron thermal velocity ($\frac{m T_i}{M T_e} \ll 1$), then the time rate of change of the electron energy can be well approximated by}:
\begin{subequations}
\label{eq:brysk}
\begin{align}
\frac{dE_e}{dt} &= -\frac{3}{2} \frac{N_e(T_e - T_i)}{\tau^E_{ei}}, \\
\label{eq:tauei}
(\tau_{ei}^E)^{-1} &= \frac{8}{3} \frac{m}{M} \frac{Z e^4 m}{\pi \hbar^3} \frac{1}{1+e^{-\mu  /T_e}} \ln \varLambda_{ei}.
\end{align}
\end{subequations}
Eq.~(\ref{eq:brysk}) was first derived by Brysk in 1974 \cite{brysk_electron-ion_1974}. In the classical limit, the relaxation time $\tau_{ei}^E$ approaches the Spitzer form ($\tau_{ei}^E \propto T_e^{-3/2}$). In the opposite limit, it becomes a constant. 

\begin{figure}
\begin{subfigure}[t]{0.5\textwidth}
\includegraphics[scale=.9]{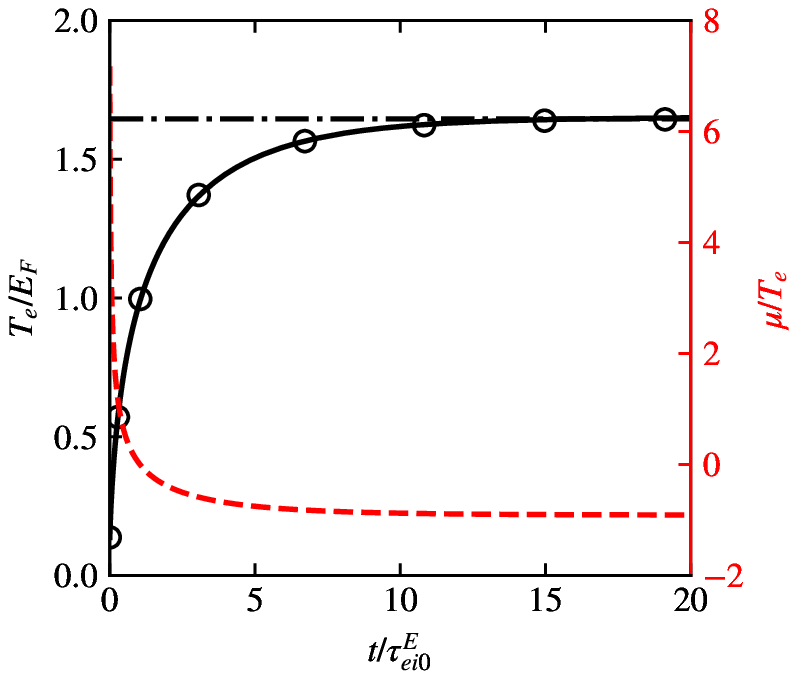}
\caption{}
\label{fig:heat_eia}
\end{subfigure}
~
\begin{subfigure}[t]{0.5\textwidth}
\includegraphics[scale=.9]{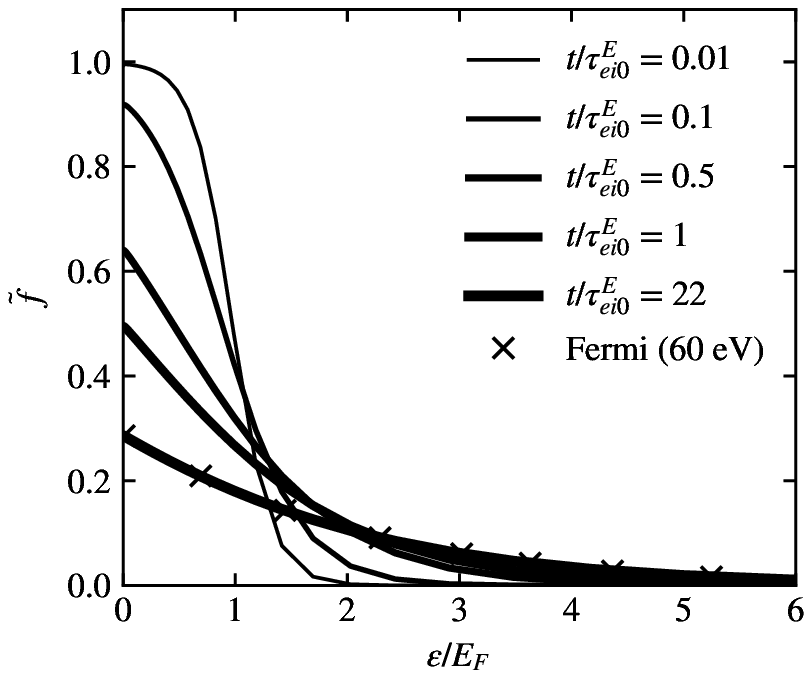}
\caption{}
\label{fig:heat_eib}
\end{subfigure}
\caption{Evolution of (a) $T_e$ and $\mu$ and (b) EEDF during heating test. The symbols in (a) are solutions obtained from Brysk's model (Eq.~(\ref{eq:brysk})), while the dashed-doted line indicates the equilibrium temperature ($T_i=60$ eV). The symbols in (b) correspond to an exact Fermi-Dirac distribution at equilibirum temperature.}
\label{fig:heat_ei}
\end{figure}

For the heating test, the initial electron distribution is a Fermi-Dirac distribution at $N_e=10^{24}$ cm$^{-3}$ and $T_e=5$ eV ($E_F = 36.5$ eV), which corresponds to $\mu/T_e=7.2$. For the ions, we set $N_i=N_e$ and $T_i = 60$ eV. Since the electron distribution is dynamically evolved, the corresponding ei collision time also changes as a function of time. For convenience, we define a reference time based on the ei collision time evaluated at initial time $\tau^E_{ei0} = \tau^E_{ei}(t=0)$, and further assume that $\ln \varLambda_{ei} = \ln \varLambda_{ee}$. Figs.~\ref{fig:heat_eia} and \ref{fig:heat_eib} show the evolution of $T_e$, $\mu$ and the EEDF during the heating process. It is evident that the temperature evolution from the qFP model (solid line) agrees very well with Brysk's model (symbols). This is expected since for Hydrogen, {$\tau_{ee} / \tau^E_{ei} \propto m/M \ll 1$}, so the instantaneous thermalization approximation is valid. Fig.~\ref{fig:heat_eib} shows that the distribution is evolved from a degenerate toward a classical distribution. This is also evident from Fig.~\ref{fig:heat_eia}, where $\mu$ is decreasing in time. The steady-state distribution (thickest line in Fig.~\ref{fig:heat_eib}) matches the analytical Fermi-Dirac distribution at 60 eV (symbols). 

\begin{figure}
\begin{subfigure}[t]{0.5\textwidth}
\includegraphics[scale=.9]{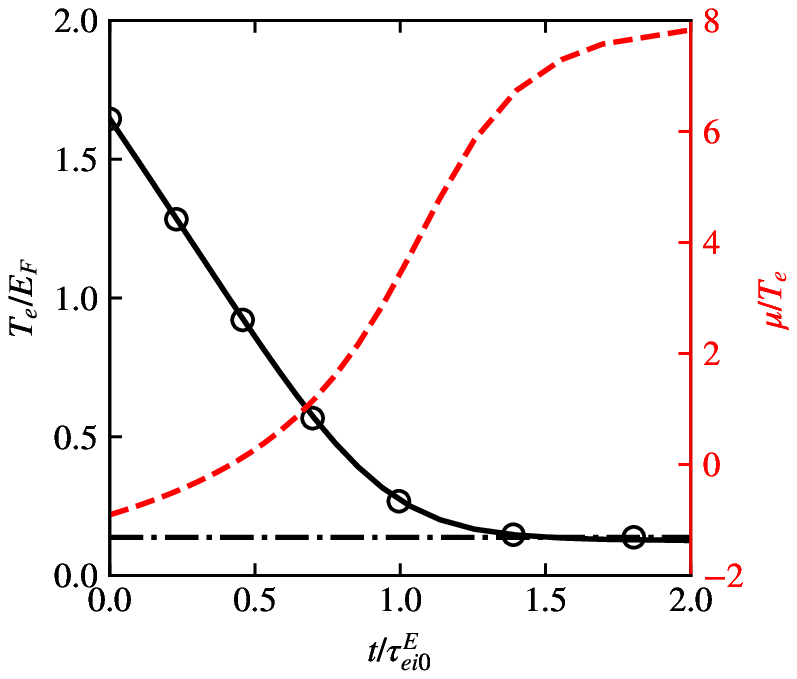}
\caption{}
\label{fig:cool_eia}
\end{subfigure}
~
\begin{subfigure}[t]{0.5\textwidth}
\includegraphics[scale=.9]{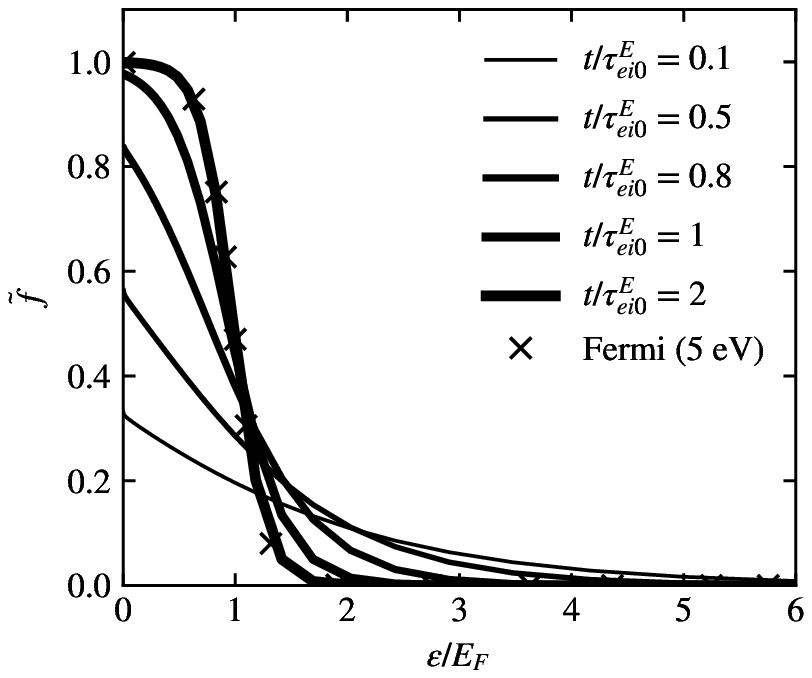}
\caption{}
\label{fig:cool_eib}
\end{subfigure}
\caption{Evolution of (a) $T_e$ and $\mu$ and (b) EEDF during cooling test. The symbols in (a) are solutions obtained from Brysk's model (Eq.~(\ref{eq:brysk})), while the dashed-doted line indicates the equilibrium temperature ($T_i=5$ eV). The symbols in (b) correspond to an exact Fermi-Dirac distribution at equilibirum temperature.}
\label{fig:cool_ei}
\end{figure}

Fot the cooling test, we initialize the electron distribution as a Fermi-Dirac distribution at $N_e=10^{24}$ cm$^{-3}$ and $T_e=60$ eV, which corresponds to $\mu/T_e=-0.9$. For the ions, we set $N_i=N_e$ and $T_i = 5$ eV. This is essentially the opposite of the heating test, where the electron distribution evolves from a classical to a degenerate distribution. Figs.~\ref{fig:cool_eia} and \ref{fig:cool_eib} show the evolution of $T_e$, $\mu$ and the EEDF during the cooling process. Similar to the heating test, the temperature evolution is in very good agreement with Brysk's results, and the distribution approaches to the correct Fermi-Dirac equilibrium at 5 eV. The steady-state distribution shows that the states below the Fermi energy are almost fully occupied, which resembles a strongly degenerate distribution. Fig.~\ref{fig:cool_ei_edist_grid} shows the steady state distributions from four different grids (16, 32, 64 and 128 groups), and indicates that the results using 64 groups are grid-independent. Although not yet converged, the results obtained from 16 and 32 groups are very reasonable, given the large energy range we need to cover to resolve the initial distribution. This highlights the advantage of nonuniform gridding capability for problems with large temperature separation.

\begin{figure}
\centering
\includegraphics[scale=.9]{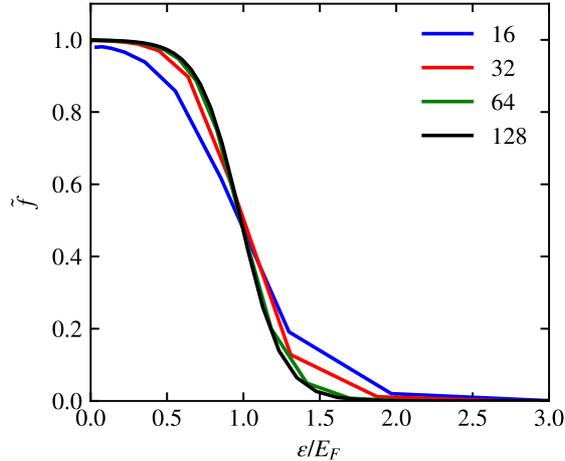}
\caption{Steady state electron energy distribution functions for the cooling problem obtained from four different energy grids (16, 32, 64 and 128 groups).}
\label{fig:cool_ei_edist_grid}
\end{figure}

\section{Conlusions}
{We construct a conservative numerical scheme for the isotropic quantum Fokker-Planck equation describing the evolution of degenerate electrons due to electron-ion and electron-electron elastic collisions. A discontinuous Galerkin method is employed to discretize the collision operators and the electron energy distribution is time integrated using a backward Euler method}. The high-order discontinuous Galerkin discretization offers additional degrees of freedom to describe energy transfer and enforce energy conservation. The fully implicit time integration relaxes the restriction on the time step, making it feasible to include other physics occurring at the slower time scales. The extension to include inelastic collisions is planned for future work. We anticipate that this capability will be useful for modeling electron collisions in high energy density physics experiments.

\section*{Acknowledgement}
I would like to thank the anonymous referees whose comments/suggestions helped improve and clarify this manuscript. I wish to thank H.A. Scott for many helpful discussions. This work was performed under the auspices of the U.S. Department of Energy by Lawrence Livermore National Laboratory under Contract DE-AC52-07NA27344.

This document was prepared as an account of work sponsored by an agency of the United States government. Neither the United States government nor Lawrence Livermore National Security, LLC, nor any of their employees makes any warranty, expressed or implied, or assumes any legal liability or responsibility for the accuracy, completeness, or usefulness of any information, apparatus, product, or process disclosed, or represents that its use would not infringe privately owned rights. Reference herein to any specific commercial product, process, or service by trade name, trademark, manufacturer, or otherwise does not necessarily constitute or imply its endorsement, recommendation, or favoring by the United States government or Lawrence Livermore National Security, LLC. The views and opinions of authors expressed herein do not necessarily state or reflect those of the United States government or Lawrence Livermore National Security, LLC, and shall not be used for advertising or product endorsement purposes.

\appendix
\section{Collision Operators}
\label{sec:qfp}
In this appendix, we formulate the ei and ee collision operators for an isotropic EEDF. Following the work of Brown and Haines \cite{brown_transport_1997}, the qFP equation for an isotropic electron velocity distribution $f_v (v)$ due to collisions with another particle distribution $F_v(V)$ (either electron or ion) can be written as:
\begin{align}
\label{eq:qfpv}
\frac{1}{Y} \frac{\partial f_{v}}{\partial t} = \frac{1}{3v^2} \frac{\partial}{\partial v} \left[ \frac{3m}{M} f_v (1- \tilde{f}_{v}) I_0 + v\left( I_2 - I'_2 + J_{-1} - J'_{-1} \right) \frac{\partial f_v}{\partial v} \right],
\end{align}
where $Y=4\pi \frac{Z^2 e^4}{m^2} \ln \varLambda$. Here $Z$ and $M$ denote charge and mass of the colliding partner, and $\ln \varLambda$ is written generically for both ei and ee collisions. Also $\tilde{f}_v \equiv \frac{h^3}{m^3 g} f_v$ is the mean occupation number where $g$ is the spin degeneracy. The set of integrals in Eq.~(\ref{eq:qfpv}) are defined as follows:
\begin{subequations}
\label{eq:integrals}
\begin{align}
I_p &= \frac{4\pi}{v^p} \int_0^v V^{p+2} F_v (V) \, dV, \quad
I_p' = \frac{4\pi}{v^p} \int_0^v V^{p+2} F_v (V) \tilde{F}_v (V) \, dV, \\
J_p &= \frac{4\pi}{v^p} \int_v^\infty V^{p+2} F_v (V) \, dV, \quad
J_p' = \frac{4\pi}{v^p} \int_v^\infty V^{p+2} F_v (V) \tilde{F}_v (V) \, dV.
\end{align}
\end{subequations}
The velocity distribution $f_v (v)$ is related to the EEDF $f_\varepsilon (\varepsilon)$ via the change of variable $f_\varepsilon (\varepsilon) d\varepsilon = 4\pi v^2 f_v (v) dv$ ($\varepsilon = mv^2/2$). Applying this to Eq.~(\ref{eq:qfpv}) and after some manipulations, we arrive at:
\begin{align}
\label{eq:qfpe}
\frac{\partial f_{\varepsilon}}{\partial t} = \frac{  m^{3/2} Y}{3 \sqrt{2}} \frac{\partial}{\partial \varepsilon} \left[ \frac{3m}{M} \frac{f_\varepsilon}{\sqrt{\varepsilon}} (1- \tilde{f}_{\varepsilon}) I_0 + 2 \varepsilon \left( I_2 - I'_2 + J_{-1} - J'_{-1} \right) \frac{\partial}{\partial \varepsilon}\left(\frac{f_\varepsilon}{\sqrt{\varepsilon}} \right)  \right]
\end{align}
where $\tilde{f}_\varepsilon = \tilde{f}_v$.

Throughout this work the ions are assumed to be non-degenerate and thermalized. If we make another assumption that the ion thermal velocity is much smaller than the electron velocity, then the set of integrals in (\ref{eq:integrals}) for ei collisions reduces to:
\begin{subequations}
\begin{align}
I_0 &\rightarrow N_i, \\
I_2 &\rightarrow \frac{3N_i T_i}{Mv^2}, \\
J_{-1} &\rightarrow 0, \\
I_2',J_{-1}' &\rightarrow 0,
\end{align}
\end{subequations}
where $N_i$ and $T_i$ are the ion density and temperature.

Inserting these results into Eq.~(\ref{eq:qfpe}), the ei collision operator for the EEDF can be written as:
\begin{align}
\left( \frac{\partial f_\varepsilon}{\partial t} \right)_{ei} =  \frac{\partial}{\partial \varepsilon} \gamma_{ei}  \left[ \frac{f_\varepsilon}{\sqrt{\varepsilon}} \left( 1 - \tilde{f}_\varepsilon \right) + T_i \frac{\partial}{\partial \varepsilon}\left(\frac{f_\varepsilon}{\sqrt{\varepsilon}} \right) \right],
\end{align}
where $\gamma_{ei} = \frac{m}{M} \sqrt{\frac{2}{m}} 2 \pi N_e Z e^4 \ln \varLambda_{ei}$. Here we also apply the charge neutrality condition, i.e., $N_e = ZN_i$.

For ee collisions, we first rewrite the integrals (\ref{eq:integrals}) in term of $f_\varepsilon$:
\begin{align}
I_0 &= 4\pi \int_0^v f_v v'^{2} \, dv' = \int_0^\varepsilon f_\varepsilon \, d \varepsilon', \\
I_2 - I_2' &= \frac{4\pi}{v^2} \int_0^v f_v \left( 1 - \tilde{f}_v \right) v'^{4} \, dv' = \varepsilon^{-1} \int_0^\varepsilon \varepsilon' f_\varepsilon \left( 1- \tilde{f}_\varepsilon \right) \, d \varepsilon', \\
J_{-1} - J_{-1}' &= \frac{4\pi}{v^{-1}} \int_v^\infty f_v \left( 1 - \tilde{f}_v \right) v' \, dv' = \varepsilon^{1/2} \int_\varepsilon^\infty \varepsilon'^{-1/2} f_\varepsilon \left( 1- \tilde{f}_\varepsilon \right) \, d \varepsilon'.
\end{align}
Let us define the following set of integrals:
\begin{subequations}
\begin{align}
L &= 3 I_0,\\
K &= 2\varepsilon \left( I_2 - I_2' + J_{-1} - J_{-1}' \right) .
\end{align}
\end{subequations}
The ee collision operator can then be written as:
\begin{align}
\left( \frac{\partial f_\varepsilon}{\partial t} \right)_{ee} = \gamma \frac{\partial}{\partial \varepsilon} \left[L \frac{f_\varepsilon}{\sqrt{\varepsilon}} (1- \tilde{f}_{\varepsilon}) + K \frac{\partial}{\partial \varepsilon}  \left( \frac{f_\varepsilon}{\sqrt{\varepsilon}} \right) \right],
\end{align}
where $\gamma = \sqrt{\frac{2}{m}} \frac{2\pi e^4 \ln \varLambda_{ee}}{3}$. For ease of notation, in the main text we drop the subscript $\varepsilon$ and consistently refer to $f$ as the electron energy distribution function. 

{
\section{Fast Evaluation of Electron-electron Flux Coefficients $K$ and $L$}
\label{app:kl}
In this appendix, we give discrete expressions of the coefficients $K_{i+1/2}$ and $L_{i+1/2}$ appeared in the expression of the ee collision fluxes (Eq. (\ref{eq:jee})), and describe an efficient method to evaluate them during the computation. The fact that these coefficients are non-local gives rise to the dense Jacobian matrix in Eq.~(\ref{eq:be2}). Analytical expressions of their derivatives with respect to the solution vector are also shown here since they are required for constructing the Jacobian matrix. For convenience, let us define the following group-wise moment of order $k$:
\begin{align}
\label{eq:mk}
M^{(k)}_i = \int_{\varOmega_i}  f (1-\tilde{f}) \varepsilon'^k \, d\varepsilon'.
\end{align}
The evaluation of Eq.~(\ref{eq:mk}) with the Pauli blocking factor is nontrivial. A practical way is to approximate the integral using Gaussian quadrature:
\begin{align}
M^{(k)}_i \simeq \frac{\Delta_i}{2} \sum_{n[i]} \sum_{p=0}^{p_{\max}} w_n \hat{f}_{i,p} \varepsilon_{i,n}^k U_p (\varepsilon_{i,n}) \left[ 1 - C \sum_{q=0}^{p_{\max}} \frac{\hat{f}_{i,q}}{\sqrt{\varepsilon_{i,n}}} U_q (\varepsilon_{i,n}) \right],
\end{align}
where $C = \frac{1}{4\pi} \left( \frac{h^2}{2m} \right)^{3/2}$. Here $n[i]$ denotes the quadrature point $n$ with weight $w_n$ and energy $\varepsilon_{i,n}$ lying within the interval $[\varepsilon_{i-1/2},\varepsilon_{i+1/2}]$. We point out that Eq.~(\ref{eq:mk}) does not need to be reevaluated at every computational cycle. Instead, for each group $i$ we can define two sets of coefficient $S_{i,p}^{(k)}$ and $T_{i,pq}^{(k)}$ $(p,q = 0, \cdots, p_{\max})$:
\begin{subequations}
\begin{align}
S_{i,p}^{(k)} &= \frac{\Delta_i}{2} \sum_{n[i]} w_n \varepsilon_{i,n}^k U_p (\varepsilon_{i,n}), \\
T_{i,pq}^{(k)} &= C \frac{\Delta_i}{2} \sum_{n[i]} w_n \varepsilon_{i,n}^{k-1/2} U_p (\varepsilon_{i,n}) U_q (\varepsilon_{i,n}).
\end{align}
\end{subequations}
Note that these coefficients only need to be calculated once at the beginning. During the calculation, we can utilize these coefficients to update the moment:
\begin{align}
M^{(k)} = \sum_{p=0}^{p_{\max}} \left( S_p^{(k)} \hat{f}_{p} - \sum_{q=0}^{p_{\max}} T_{pq}^{(k)} \hat{f}_{p} \hat{f}_{q} \right).
\end{align}
Here we have omitted the group index for ease of notation. The derivatives of $M^{(k)}$ with respect to the expansion coefficient $\hat{f}_{p}$ can be evaluated as follows:
\begin{align}
\frac{ \partial M^{(k)} }{\partial \hat{f}_{p}} = S_p^{(k)} - \sum_{q=0}^{p_{\max}} (1 + \delta_{pq}) T_{pq}^{(k)} \hat{f}_{q}.
\end{align}

Using (\ref{eq:mk}) and (\ref{eq:ni}), the coefficients $K_{i+1/2}$ and $L_{i+1/2}$ defined in Eq.~(\ref{eq:ee}) can now be written as: 
\begin{subequations}
\begin{align}
K_{i+1/2} &= 2 \sum_{j=1}^i  M^{(1)}_j  + 2 (\varepsilon_{i+1/2})^{3/2} \sum_{j=i+1}^{N} M^{(1/2)}_j,  \\
L_{i+1/2} &= \frac{3}{\sqrt{2}} \sum_{j=1}^i \Delta_j \hat{f}_{i,0}.
\end{align}
\end{subequations}
The derivatives of $K_{i+1/2}$ and $L_{i+1/2}$ with respect to a group coefficient $\hat{f}_{j,p}$ can be evaluated as:
\begin{subequations}
\begin{align}
\frac{ \partial K_{i+1/2}}{\partial \hat{f}_{j,p} } & = 
\begin{cases}
2 \frac{\partial M^{(1)}_j}{\partial \hat{f}_{j,p}}, & \text{for}\ j \le i  \\
2 (\varepsilon_{i+1/2})^{3/2} \frac{\partial M^{(1/2)}_j}{\partial \hat{f}_{j,p}}, & \text{otherwise}
\end{cases} \\
\frac{ \partial L_{i+1/2}}{\partial \hat{f}_{j,p} } & = 
\begin{cases}
\delta_{p0} \frac{3}{\sqrt{2}} \Delta_j , & \text{for}\ j \le i  \\
0, & \text{otherwise}
\end{cases}
\end{align}
\end{subequations}
We remark that since $K_{i+1/2}$ and $L_{i+1/2}$ depend on solution of all energy groups, the Jacobian containing the derivatives $\frac{ \partial K_{i+1/2}}{\partial \hat{f}_{j,p} }$ and $\frac{ \partial L_{i+1/2}}{\partial \hat{f}_{j,p} }$ is a dense matrix.

\section{Energy grid generation}
\label{app:grid}
We define an energy grid that has the grid spacings increased by a constant ratio $r$:
\begin{equation}
\varepsilon_{i+1/2} = \varepsilon_{i-1/2} + r^{i-1} \Delta \varepsilon \ \text{for} \ i=1,2,\cdots,N_\varepsilon
\label{eq:gridgen}
\end{equation}
where $N_\varepsilon$ is the number of energy groups. By recursively applying (\ref{eq:gridgen}), we obtain an expression relating the first and last grid points:
\begin{equation}
\varepsilon_{N+1/2} = \varepsilon_{1/2} + \Delta \varepsilon \sum_{i=0}^{N_\varepsilon-1} r^{i}
\end{equation}
By rewriting $\sum_{i=0}^{N_\varepsilon-1} r^{i-1} = (1-r^{N_\varepsilon}) \sum_{i=0}^{\infty} r^{i}  = (1-r^{N_\varepsilon})/(1-r) $, we obtain an expression for determining $\Delta \varepsilon$:
\begin{equation}
\Delta \varepsilon = \frac{(\varepsilon_{N_\varepsilon+1/2} - \varepsilon_{1/2})(1-r)}{1-r^N_\varepsilon}
\end{equation}
Hence for a energy grid consisting of $N_\varepsilon$ groups spanning from $\varepsilon_{\min}$ to $\varepsilon_{\max}$, we can estimate the initial grid spacing $\Delta \varepsilon$ by:
\begin{equation}
\label{eq:egrid}
\Delta \varepsilon = \frac{(\varepsilon_{\max} - \varepsilon_{\min})(1-r)}{1-r^{N_\varepsilon}}
\end{equation}
In practice, we choose $\varepsilon_{\min}$ to be a small but non-zero quantity to retain resolution in the low energy portion of the spectrum. However, the physical domain should always start from zero, so one can either add an zero energy grid point or simply set the first value of the energy grid to zero.
}

\end{document}